\documentclass[journal=jacsat,manuscript=article]{achemso}

\usepackage[version=3]{mhchem} 
\usepackage{appendix}
\usepackage{xcolor}
\usepackage{url}

\author{Tian-Yi Zhang}
\affiliation{International Center for Quantum Materials, School of Physics, Peking University, Beijing 100871, China}

\author{Yue Mao}
\affiliation{International Center for Quantum Materials, School of Physics, Peking University, Beijing 100871, China}

\author{Peng-Yi Liu}
\affiliation{International Center for Quantum Materials, School of Physics, Peking University, Beijing 100871, China}

\author{Ai-Min Guo}
\affiliation{Hunan Key Laboratory for Super-microstructure and Ultrafast Process, School of Physics, Central South University, Changsha 410083, China
}

\author{Qing-Feng Sun}
\email{sunqf@pku.edu.cn}
\affiliation{International Center for Quantum Materials, School of Physics, Peking University, Beijing 100871, China}
\affiliation{Hefei National Laboratory, Hefei 230088, China
}

\title[An \textsf{achemso} demo]
  {Anomalous Magnetoresistance beyond the Jullière Model for Spin Selectivity in Chiral Molecules}

\keywords{American Chemical Society, \LaTeX}

\begin{tocentry}
\centering
\includegraphics{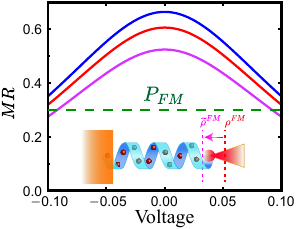}
\end{tocentry}

\begin{document}


\begin{abstract}
The issue of anomalous high magnetoresistance, beyond the Jullière model, observed in nonmagnetic electrode-chiral molecular-ferromagnetic electrode devices has puzzled the community for a long time.
Here, by considering the magnetic proximity effect which shifts the nonmagnetic-ferromagnetic interface toward chiral molecules, we show the anomalous high magnetoresistance beyond the spin polarization in ferromagnetic electrodes even in the very weak spin-orbit coupling.
Our results are in excellent agreement with the experiments, demonstrating that the spin-orbit coupling plays a fundamental role in chiral-induced spin selectivity and the magnetic proximity effect can dramatically enhance the magnetoresistance.
These results elucidate the interaction between chiral molecules and ferromagnetic electrodes and facilitate the design of chiral-based spintronic devices.
\end{abstract}


Chirality plays a fundamental role in nature and has attracted extensive interest among the chemistry, biology, and physics communities.\citep{ref1,ref2} In the last decade, a lot of work has studied the spin transport along chiral molecules,\citep{ref3,ref4} finding that spin-unpolarized electrons will become highly spin-polarized when transmitted through chiral molecules, namely the chiral-induced spin selectivity (CISS).\citep{ref5,ref6,ref7} This CISS holds great applications in spintronic devices. Historically, CISS was initially observed in stearoyl lysine photoemission experiments in 1999.\citep{ref8} Since then, CISS has been reported in different chiral materials,\citep{ref9,ref_natmat2024,ref_sciadv2025} such as double-stranded DNA,\citep{ref3,ref10} $\alpha$-helical proteins,\citep{ref11} and halide perovskites.\citep{ref12} To understand the physical mechanism behind CISS, many theories have been proposed.\citep{ref13,ref14,ref15,refPRL,refPRB_ZTY,refJPCL_LPY,ref16,ref18,ref19,ref20,ref_Isr_JF,ref_revJCP_2023} The first kind of theory suggests that the helical structure and the spin-orbit coupling (SOC) of chiral molecules are key to generating large spin polarization for electrons passing through chiral molecules.\citep{ref13,ref14,ref15}
In these works, the SOC strength is usually set to be about one-tenth of the inter-nucleobase or inter-amino acid hopping integral.
The second kind of theory considers that CISS arises from the interaction between metal substrates with large SOC and chiral molecules.\citep{ref16} However, recent experiments demonstrate that CISS still exists in the absence of metal substrates, revealing substrates with large SOC are not necessary for CISS.\citep{ref17,ref_JACS_hole,ref_PNAS_hole} Although many other models such as spin-dependent scattering,\citep{ref18} electron-phonon coupling,\citep{ref19} and electron correlation \citep{ref20} are proposed, the physical mechanism of CISS remains under debate.

\begin{figure}
\centering
\includegraphics[width = 0.9 \linewidth]{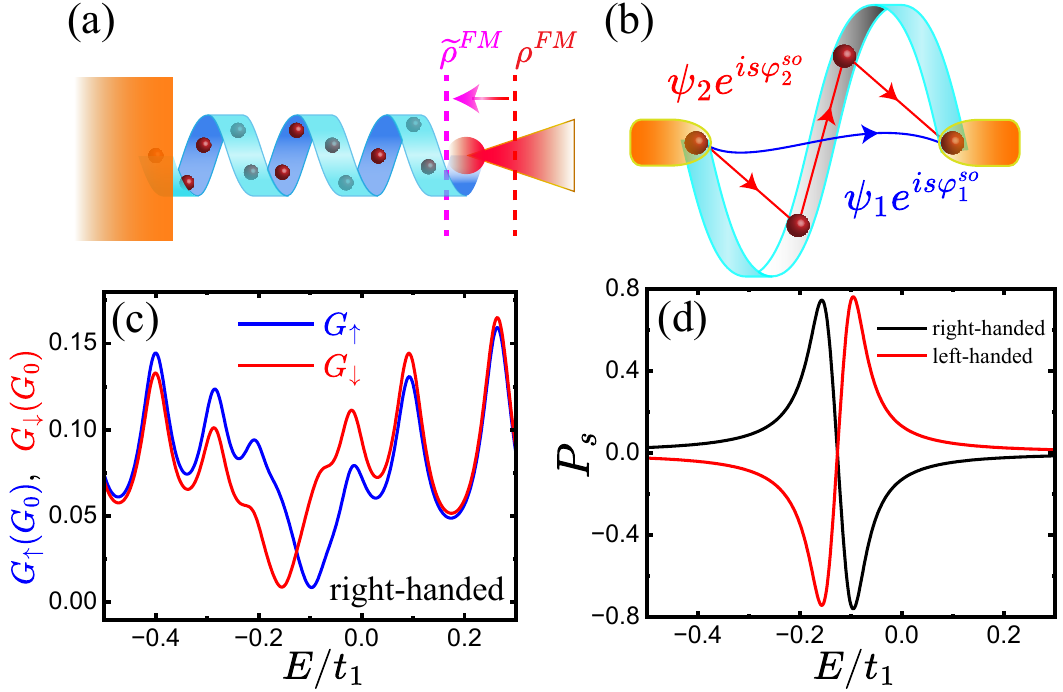}
\caption{\label{fig1} Schematic of chiral molecular model and the corresponding spin polarization.
 (a) Schematic of a N-chiral molecule-FM (or N) device.
 The right electrode is set to be nonmagnetic to study the spin polarization $P_s$ for electron propagation through chiral molecules, or to be ferromagnetic to study the MR. In the presence of MPE induced by the FM electrode, the rightmost site of chiral molecules becomes magnetized.
 (b) Schematic diagram of two-pathway interference.
 (c) and (d) are conductances $G_\uparrow$, $G_\downarrow$ and spin polarization $P_s$ versus the energy $E$ in N-chiral molecule-N device at extremely weak SOC $s_1=0.01t_1$. Here, $E$ is the Fermi energy, $t_1$ is the intrachain hopping integral and $G_0=e^2/h$ is the quantum conductance.}
\end{figure}

In experiments, a variety of techniques have been employed to measure the CISS effect, including photoemission,\citep{ref3} electrochemistry,\citep{ref21} fluorescence signals,\citep{ref22} and electrical transport,\citep{ref23,ref24} where the electrical transport method is extremely powerful for detecting the CISS effect.
In electrical transport experiments, CISS is studied by measuring the magnetoresistance (MR) of a two-terminal
nonmagnetic electrode-chiral molecule-ferromagnetic electrode (N-chiral molecule-FM)
device,\citep{ref25,ref26} as shown in Figure \ref{fig1}a.
When the magnetization direction of the FM electrode is oriented in positive or negative directions, the current flowing through the device differs, and one can obtain the MR and thus determine the spin selectivity of chiral molecules.\citep{ref27,ref28} Provided that the spin-up and spin-down density of states (DOSs) in FM electrodes are $\rho_{\uparrow}^{FM}$ and $\rho_{\downarrow}^{FM}$, respectively, the FM polarization is defined as \citep{ref29} $P_{FM}=(\rho_{\uparrow}^{FM}-\rho_{\downarrow}^{FM})/(\rho_{\uparrow}^{FM}+\rho_{\downarrow}^{FM})$. Similarly, the spin polarization evaluating the CISS effect of chiral molecules is defined as $P_s=(n_{\uparrow}^{mol}-n_{\downarrow}^{mol})/(n_{\uparrow}^{mol}+n_{\downarrow}^{mol})$, where $n_{\uparrow}^{mol}$ and $n_{\downarrow}^{mol}$ are, respectively, the densities of spin-up and spin-down electrons passing through chiral molecules. According to the Jullière model,\citep{ref30} the current $I_{+M}$ under positive magnetization of FM electrodes is proportional to $n_{\uparrow}^{mol}\rho_{\uparrow}^{FM}+n_{\downarrow}^{mol}\rho_{\downarrow}^{FM}$. By reversing the magnetization, the spin-up and spin-down DOSs in FM electrodes exchange, and the current under negative magnetization satisfies $I_{-M}\propto{n_\uparrow^{mol}\rho}_\downarrow^{FM}+n_\downarrow^{mol}\rho_\uparrow^{FM}$. The experimentally measured MR can then be expressed as $MR=(I_{+M}-I_{-M})/(I_{+M}+I_{-M})$, which is equal to $P_s P_{FM}$. Accordingly, the MR is always smaller than the FM polarization $P_{FM}$ as $|P_s|\le 1$. FM electrodes used in experiments are usually composed of transition metals such as nickel (Ni), iron (Fe), and cobalt (Co), with $P_{FM}$ about 33\%, 44\%, and 45\%, respectively.\citep{ref31} However, it has been widely reported that the MR associated with the CISS effect can reach 80\%,\citep{ref32,ref33,ref34} which is much larger than $P_{FM}$. This violation of the Jullière model frequently appears in electrical transport experiments of the CISS and has been noticed for a long time.\citep{ref35,ref36,ref37} Many researchers find it incomprehensible and feel puzzled, but a satisfactory answer to this question remains elusive to this day.\citep{ref38}

In this Letter, we study the CISS in chiral molecules and demonstrate strong CISS even at extremely weak SOC regime.
When the magnetic proximity effect (MPE) is introduced at the molecular end, an anomalous large MR far beyond FM polarization $P_{FM}$ is found.
We show that the MR is always lower than $P_{FM}$ in the absence of MPE, and the MR completely disappears once either chirality or SOC vanishes, revealing their essential roles in the CISS and MPE is the origin of $MR>P_{FM}$. These findings well address the dilemmas mentioned above.

We first show the origin of the CISS effect in the nonmagnetic electrode-chiral molecule-nonmagnetic electrode (N-chiral molecule-N) device. The physical origin of CISS arises from the multi-pathway helical structure and SOC of chiral molecules.\citep{ref13,ref14,ref39} There exist multiple pathways for electron propagation through chiral molecules, and the interference of electron wave-functions through these pathways is the basis for the CISS effect. If such interference is absent, CISS effect cannot occur in one-pathway molecules.\citep{ref13}
Here, we consider the simplest case of two transport pathways as an example, as illustrated in Figure \ref{fig1}b, where $\psi_1$ and $\psi_2$ describe the wave-functions of electron propagation along pathway-1 and pathway-2, respectively, in the absence of SOC. Due to the chiral structure, the two pathways are usually different from each other with the phase difference being $\Delta \varphi_{12}$. In the presence of SOC, the two wave-functions are changed to $\psi_1e^{is\varphi_1^{so}}$ and $\psi_2e^{is\varphi_2^{so}}$, generating the spin-related phase factors.\citep{ref40,ref_PRB_QD} Here, $s= \uparrow(+)$ and $\downarrow(-)$ denote the up and down spin, respectively, and $\varphi_{1/2}^{so}$ are the phase related to the SOC. The total probability of electron propagation through the two pathways is proportional to${\ \left|\psi_1e^{is\varphi_1^{so}}+\psi_2e^{is\varphi_2^{so}}\right|}^2=\left|\psi_1\right|^2+\left|\psi_2\right|^2+2\left|\psi_1\right|\left|\psi_2\right|\cos (s\Delta \varphi_{so}+\Delta \varphi_{12})$ with $\Delta \varphi_{so}=\varphi_1^{so}-\varphi_2^{so}$, which depends on the electron spin. As a result, the CISS emerges. In realistic systems, there may be more than two interference pathways, and the magnitude of CISS effect is determined by the details of the molecule.

Let us study the spin transport properties of the N-chiral molecule-N device (see Figure \ref{fig1}a).
We take the single-helical protein as an example,\citep{ref14} which can be described by a tight-binding Hamiltonian with SOC and multi-pathway structure:
\begin{equation}
  \begin{aligned}
    H_{cmolecule}=H_{mol}+H_{so} &=\sum_{n=1}^N \varepsilon_n c_n^\dagger c_n +\sum_{n=1}^{N-1} \sum_{j=1}^{N-n} [t_j c_n^\dagger c_{n+j}+h.c.]\\
    &+\sum_{n=1}^{N-1} \sum_{j=1}^{N-n} [2 i s_j \cos(\varphi_{n,j}^{-})c_n^\dagger \sigma_{nj} c_{n+j} + h.c.]
  \end{aligned}
  \label{eq_Hami}
\end{equation}
where $c_n^\dagger = (c_{n,\uparrow}^\dagger,c_{n,\downarrow}^\dagger)$ is the creation operator at site $n$, and the molecular length is $N$.
$\varepsilon_n$ is the on-site energy, $t_j$ is the long-range hopping integral between sites $n$ and $n+j$, and $s_j$ is the corresponding SOC.
$\sigma_{nj}$ is a $2 \times 2$ matrix related to the molecular structure.
$\varphi_{n,j}^{-} = j \Delta \varphi /2$, where $\Delta \varphi$ is the twist angle between two neighboring sites.
The length of chiral molecule is $N=30$.
The twist angle is set to $\Delta \varphi=5\pi/9$ for right-handed molecules and $\Delta \varphi=-5\pi/9$ for left-handed molecules (see Section S1.1, S1.2, and Figure S1 in Supporting Information for the Hamiltonian, parameters, and multi-pathway structure). The robust explanatory power of this model is demonstrated by its successful explanation of key CISS-related phenomena, including the CISS effect in chiral molecules,\citep{ref15}
the dynamic process of the CISS effect in electron donor--chiral molecule--acceptor systems,\citep{refPRB_ZTY} the spin polarization and magnetization effect in chiral molecules,\citep{refJPCL_LPY} CISS-modulated spin-to-charge conversion,\citep{refJPCL_liu2025} and inverse CISS effect.\citep{ref_zhangICISS}
The structural parameters are identical to previous works,\citep{ref14,ref15} except that the SOC strength $s_1$ is set to be two orders of magnitude smaller than the intrachain hopping integral $t_1$. Setting the hopping integral $t_1 = 100$ meV,\citep{ref47} the SOC strength is only $s_1 = 1$ meV, which is consistent with the actual values in the experiments.\citep{ref6,ref9,ref48} Below we will show that this extremely weak SOC is sufficient to produce large spin polarization.

The conductance can be calculated using the Landauer-Büttiker formalism (see Section S1.3, S1.4 in the Supporting Information),\citep{ref_XYX_PRB} and the spin polarization is $P_s=(n_\uparrow^{mol}-n_\downarrow^{mol})/(n_\uparrow^{mol}+n_\downarrow^{mol})=(G_\uparrow-G_\downarrow)/(G_\uparrow+G_\downarrow)$, where $G_\uparrow$ and $G_\downarrow$ correspond to the spin-up and spin-down conductances.
Figure \ref{fig1}c shows $G_\uparrow$ and $G_\downarrow$ vs. Fermi energy $E$. One can see that $G_\uparrow$ is different from $G_\downarrow$ in a wide energy range. This indicates spin-unpolarized incident electrons from the left nonmagnetic electrode become highly spin-polarized when they transmit through chiral molecules to the right nonmagnetic electrode, a clear signature of CISS.
Figure \ref{fig1}d shows the spin polarization $P_s$ derived from Figure \ref{fig1}c. $P_s$ can reach the value of 74.4\% even at the extremely weak SOC with $s_1=0.01t_1$.
The physical origin of CISS arises from the combination of SOC and multi-pathway interference, as shown in Figure \ref{fig1}b.

To demonstrate the significance of SOC and chirality, we calculate $G_\uparrow$, $G_\downarrow$, and $P_s$ in the absence of either SOC or chirality
and analyse the phyical picture from the multi-pathway interference
(see Section S1.5, Figures S2-S4, and Table S1 in the Supporting Information). In both cases, $G_\uparrow$ and $G_\downarrow$ completely overlap and $P_s$ is exactly zero.
Furthermore, $P_s$ in right-handed molecules are exactly opposite to that in left-handed molecules (see Figure \ref{fig1}d, Section S1.5 and Figure S2
in the Supporting Information).
This indicates that chirality is the prerequisite for CISS,
and the SOC plays a fundamental role in the CISS although it is extremely weak.
Due to $P_s$ of left-handed and right-handed molecules being exactly opposite,
we present only the results for the right-handed molecules below.

\begin{figure}
\centering
\includegraphics[width = 0.9 \linewidth]{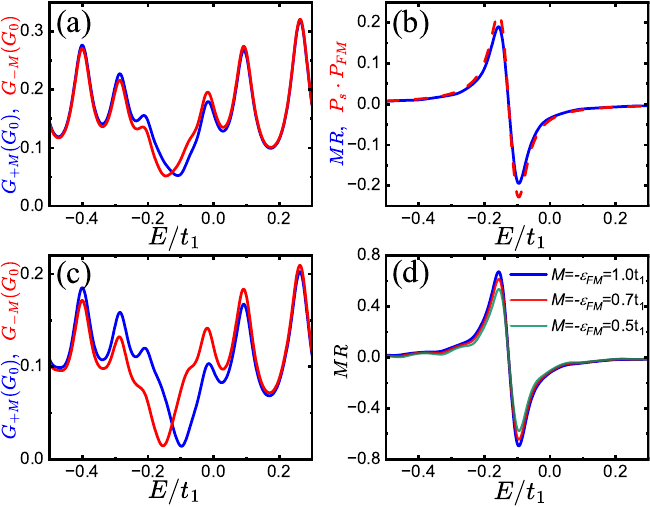}
 \caption{\label{fig2} Conductance and MR in the N-molecule-FM device.
 (a) and (b) respectively show conductances $G_{+M}$, $G_{-M}$ and MR,
 $P_s P_{FM}$ versus Fermi energy $E$ in the device without MPE.
 Here $MR\le P_s P_{FM}<P_{FM}$ implies the validity of the Jullière model.
 (c) Conductances $G_{+M}$ and $G_{-M}$ versus $E$ by considering the MPE ($M=-\varepsilon_{FM}=t_1$). (d) MR in the presence of MPE for different MPE parameters, in which MR can be larger than $P_{FM}=30\%$, implying the invalidity of the Jullière model. In (a-d), the SOC strength $s_1=0.01t_1$.
}
\end{figure}

Next, we study the MR in N-chiral molecule-FM device, see Figure \ref{fig1}a.
The FM polarization is fixed at $P_{FM}=30\%$, a typical value of the FM tip/FM substrate used in the experiments.
The MR is calculated by $MR=(G_{+M}-G_{-M})/(G_{+M}+G_{-M})$, where $G_{+M} (G_{-M})$ represents the conductance for positive (negative) magnetization electrode (see Section S1.3 in the Supporting Information).
Figure \ref{fig2}a and Figure \ref{fig2}b display, respectively, the conductance and MR without MPE. The difference between conductances $G_{+M}$ and $G_{-M}$ is too small to yield a notable MR compared to the MR observed experimentally.\citep{ref27,ref28} In Figure \ref{fig2}b we also show the product of $P_s$ and $P_{FM}$.
MR is slightly smaller than $P_s P_{FM}$ only, which is consistent with the Jullière model.
We emphasize that, for whatever the values of model parameters,
$MR<P_{FM}$ always and the Jullière model holds.
In the experiments, however, MR can be larger than $P_{FM}$ and the Jullière model is violated.\citep{ref33,ref34}

\begin{figure}
\centering
\includegraphics[width = 0.9 \linewidth]{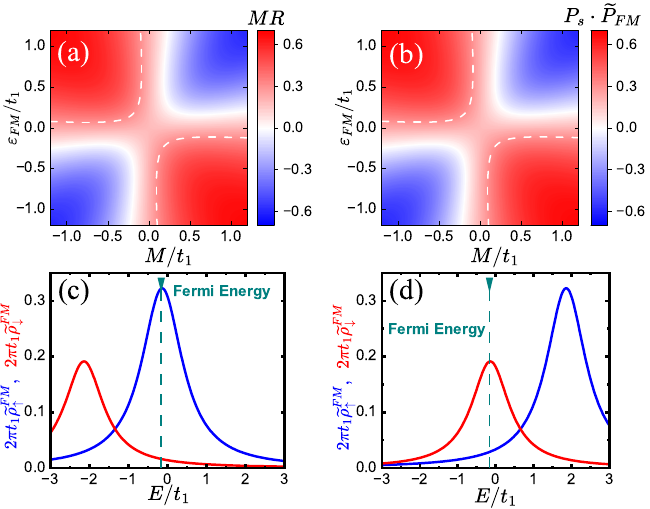}
\caption{\label{fig3} MR vs. parameters of the MPE and the physical mechanism for $MR>P_{FM}$.
(a) 2D plot of MR versus the parameters  $M$ and $\varepsilon_{FM}$ of the MPE at the Fermi level $E=-0.16t_1$. The dashed-white line represents $MR=P_{FM}=30\%$. (b) $P_s{\widetilde{P}}_{FM}$ as a function of $M$ and $\varepsilon_{FM}$. Here the $P_s{\widetilde{P}}_{FM}$ is almost identical to MR in (a). (c) and (d) Modified spin-up (blue line) and spin-down (red line) DOSs versus the energy $E$ for $M=-\varepsilon_{FM}=t_1$ (c) and $M=\varepsilon_{FM}=t_1$ (d). The green-dashed-vertical lines show $E=-0.16t_1$, the value of which is used in (a,b).
}
\end{figure}

To answer the question why $MR>P_{FM}$ in many experiments, we consider the MPE induced by the FM electrode. When two different materials are coupled together, their electronic wave-functions hybridize. Such hybridization becomes stronger at the end site of chiral molecules closest to the FM material because the discrete energy levels in the molecules are coupled to many states in the FM electrode.
This MPE can be described by the Hamiltonian $H_{MPE}=\varepsilon_{FM} c_N^\dagger c_N + Mc_N^\dagger \sigma_z c_N$, where $M$ represents the magnitude of the magnetization, $\varepsilon_{FM}$ represents the renormalization energy on the rightmost site (i.e. site $N$) closest to the FM electrode due to the MPE (see Sections S1.2 and S1.6 in the Supporting Information), and $\sigma_z$ is the Pauli matrix for the $z$-direction. The values of $M$ and $\varepsilon_{FM}$ are about hundreds of meV.\citep{refNat,refNano,refNatNano} As illustrated in Figure \ref{fig1}a, the actual system consists of a nonmagnetic substrate, a chiral molecule with the rightmost site influenced by the MPE, and a FM tip.
In the presence of MPE, the conductances and the MR can be calculated as before (see Section S1.3 in the Supporting Information).
From Figure \ref{fig2}c, a large difference between $G_{+M}$ and $G_{-M}$ is shown, which is entirely different from Figure \ref{fig2}a without the MPE.
Figure \ref{fig2}d displays the MR for different parameters of the MPE.
It is clear that MR is dramatically enhanced and can achieve the value of 67.3\%, greater than $P_{FM}=30\%$.
Moreover, with increasing the molecular length, MR also increases and can surpass 80\% (see Figure S5a), which is in excellent agreement with the experiments.\citep{ref35,ref49}
We also study the influence of dephasing and SOC \citep{ref42,ref43,ref44,ref45,ref46} on the MR, finding that high MR beyond $P_{FM}$ is achieved in a large parameter range (see Figures S5b and S6 in the Supporting Information).
To further explore the influence of the MPE, in Figure \ref{fig3}a we calculate MR versus the parameters, $M$ and $\varepsilon_{FM}$, related to the MPE, where the white-dashed line corresponds to the FM polarization $P_{FM}=30\%$. One can see that MR can be greater than $P_{FM}$, indicating the violation of the Jullière model in a wide parameter range (red region in Figure \ref{fig3}a; note that these regions are accessible since $M$ and $\varepsilon_{FM}$ can attain values of hundreds of meV,\citep{refNat,refNano,refNatNano} i.e., a few $t_1$). To maintain this, $M$ and $\varepsilon_{FM}$ should have opposite sign, and the maximum of MR occurs for $M=-\varepsilon_{FM}$. On the other hand, when $M$ and $\varepsilon_{FM}$ have the same sign, MR is reversed and its absolute value exceeds $P_{FM}$ as well (Figure \ref{fig3}a).

Let us demonstrate analytically why MR can far exceed $P_{FM}$.
The physical mechanism arises from the shift of nonmagnetic-FM interface in the presence of MPE. In the absence of MPE, the nonmagnetic-FM interface locates between the rightmost site, $N$, of chiral molecules and the FM electrode, as indicated by the red-dashed line in Figure \ref{fig1}a. According to the Jullière model, the device MR always satisfies $MR=P_s P_{FM}<P_{FM}$.
While the MPE occurs, the rightmost site is magnetized, and subsequently the nonmagnetic-FM interface is shifted to the position between the $N-1$th site and the $N$th one, see the pink-dashed line Figure \ref{fig1}a. Because of the shift of nonmagnetic-FM interface, the device MR is modified to $MR=P_s{\widetilde{P}}_{FM}$ according to the Jullière model, which depends on the spin polarization ${\widetilde{P}}_{FM}$ of the $N$th site instead of that in the FM electrode.
The modified DOS at the $N$th site is (see details in Section S1.6 in the Supporting Information):
\begin{equation}
    {\widetilde{\rho}}_s^{FM}=\frac{1}{2\pi}\frac{\Gamma_s^{FM}}{\left(E-\varepsilon_{FM}-sM-\varepsilon_0\right)^2+\left(\Gamma_s^{FM}+\gamma_0\right)^2/4},
\label{eq2}
\end{equation}
where $s=\uparrow(+)$ and $\downarrow(-)$,
the linewidth function $\Gamma_s^{FM}$ is proportional
to the DOS $\rho_s^{FM} $ of the FM electrode,
and $\varepsilon_0$, $\gamma_0$ are the parameters determined by chiral molecules
(see Section S1.6 in the Supporting Information).
Consequently, the modified FM polarization is ${\widetilde{P}}_{FM}=({\widetilde{\rho}}_\uparrow^{FM}-{\widetilde{\rho}}_\downarrow^{FM})/({\widetilde{\rho}}_\uparrow^{FM}+{\widetilde{\rho}}_\downarrow^{FM})$ of the $N$th site, which not only depends on the DOS of the FM electrode but also on the parameters, $M$ and $\varepsilon_{FM}$, of the MPE. For suitable $M$ and $\varepsilon_{FM}$, the modified DOS for specific spin may be very small, resulting in 100\% modified polarization $\widetilde{P}_{FM}$ so that $\widetilde{P}_{FM}$ will be much larger than $P_{FM}$. As a result, the device $MR=P_s \widetilde{P}_{FM}$, can be greater than $P_{FM}$ for large ${\widetilde{P}}_{FM}$.

Figure \ref{fig3}b shows the analytical result of $P_s\widetilde{P}_{FM}$ by using the fitted parameters $\varepsilon_0$, $\gamma_0$ (see details in Section S1.6 in the Supporting Information) and Eq.(\ref{eq2}). As compared with Figure \ref{fig3}a, one can see from Figure \ref{fig3}b that the analytical result $P_s\widetilde{P}_{FM}$ matches the numerical MR perfectly. This indicates that our theoretical model can quantitatively describe the influence of MPE on chiral molecules.

\begin{figure}
\centering
\includegraphics[width = 0.9 \linewidth]{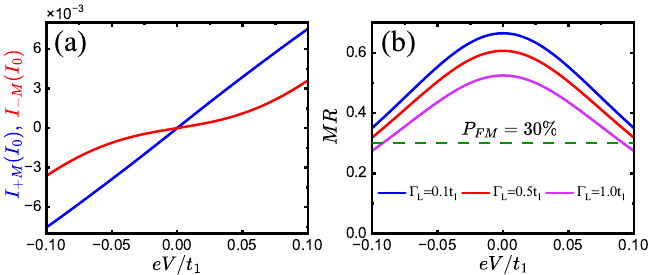}
\caption{\label{fig4} I-V results for finite bias.
(a) Current of the device versus the voltage $V$ with the right FM electrode under positive and negative magnetizations for $\Gamma_L=0.1t_1$. (b) MR versus the voltage $V$ for different $\Gamma_L$. The green dashed line is $P_{FM}=30\%$. In (a) and (b), the SOC strength $s_1=0.01t_1$, $E=-0.16t_1$ and $M={-\varepsilon}_{FM}=t_1$, and $I_0=e t_1/h$ is the current unit.}
\end{figure}

To further investigate the influence of MPE, Figure \ref{fig3}c and Figure \ref{fig3}d depict the modified spin-up and spin-down DOSs, ${\widetilde{\rho}}_\uparrow^{FM}$ and ${\widetilde{\rho}}_\downarrow^{FM}$, of the rightmost site $N$ of chiral molecules for $M=-\varepsilon_{FM}=t_1$ and $M=\varepsilon_{FM}=t_1$, respectively. One can see that ${\widetilde{\rho}}_\uparrow^{FM}$ and ${\widetilde{\rho}}_\downarrow^{FM}$ exhibit significant differences in a wide energy range. Specifically, when $M$ and $\varepsilon_{FM}$ have opposite sign, according to Eq.(\ref{eq2}), ${\widetilde{\rho}}_\uparrow^{FM}$ is much larger than ${\widetilde{\rho}}_\downarrow^{FM}$ at the Fermi energy (see the green-dashed-vertical line in Figure \ref{fig3}c). This large difference leads to greatly modified polarization ${\widetilde{P}}_{FM}$. When the magnetization parameter is increased, the spin-down DOS ${\widetilde{\rho}}_\downarrow^{FM}$ becomes much smaller and ${\widetilde{P}}_{FM}$ will be very close to 100\%. As a result, $MR\approx P_s>P_{FM}$. Contrarily, when $M$ and $\varepsilon_{FM}$ have the same sign, ${\widetilde{\rho}}_\downarrow^{FM}>{\widetilde{\rho}}_\uparrow^{FM}$ at the Fermi energy (Figure \ref{fig3}d) and ${\widetilde{P}}_{FM}$ will be reversed, leading to negative MR.

For direct comparison with the experiments, we calculate the current in the
N-chiral molecule-FM device under different bias voltages by considering the MPE. The current flowing across the device is given by\citep{ref50}
$    I_{\pm M}=\frac{1}{e}\int_{E-eV/2}^{E+eV/2}{G_{\pm M}\left(\varepsilon\right)d\varepsilon},$
with $V$ the voltage.
Then $MR=(I_{+M}-I_{-M})/(I_{+M}+I_{-M})$.
In Figure S7a in Supporting Information, we calculate the corresponding I-V curves in the absence of MPE.
The difference between $I_{+M}$ and $I_{-M}$ is too small to generate considerable MR.
Figure S7b depicts MR for different electrode-molecule coupling strengths $\Gamma_L$. The results show that MR is small, consistently lower than $P_{FM}=30\%$, exhibits a symmetric shape and a downward shift with increasing $\Gamma_L$.
The symmetry arises because $I_{\pm M}$ is an odd function of $V$, making the MR an even function of $V$.
It is worth emphasizing that in the absence of MPE, MR is always lower than $P_{FM}$, no matter how we change the model parameters, including the SOC strength, the dephasing strength, the electrode-molecule coupling, and the molecular parameters.

The situation completely changes when we consider the MPE.
Figure \ref{fig4}a illustrates the currents $I_{+M}$ and $I_{-M}$ versus the voltage $V$ in the presence of MPE.
By increasing $V$, the I-V curves are nonlinear and exhibit antisymmetric behavior. Notably, the difference between $I_{+M}$ and $I_{-M}$ is significantly large over a wide range of voltage, leading to very large MR. Figure \ref{fig4}b shows MR versus the voltage $V$, which exhibits a symmetric shape and a downward shift with increasing $\Gamma_L$.
The symmetry arises because $I_{\pm M}$ is an odd function of $V$, rendering the MR an even function of $V$. The downward shift can be attributed to the decrease in ${\widetilde{P}}_{FM}$ with increasing $\Gamma_L$, as illustrated in Figure S8.
It is evident that MR can be much greater than $P_{FM}$, which is in excellent agreement with the experiments \citep{ref27,ref28,ref33,ref49} and violates the Jullière model.
The maximum of MR exceeds 60\%, which is twice as much as $P_{FM}$. Furthermore, $MR>P_{FM}$ can occur in a wide range of parameters, although the SOC is extremely weak.

Finally, in the absence of SOC, $s_1=0$, the MR is exactly zero (see Figure S9 in the Supporting Information), and the CISS disappears, no matter how we change the structural parameters or whether the MPE is considered. This means that the extremely weak SOC plays a pivotal role for the CISS and the MPE is the determining factor for $MR>P_{FM}$.
Also, $MR>P_{FM}$ remains even in the presence of significant twist angle disorder as long as the MPE is included (see Figure S10 in the Supporting Information). Conductance dependence on FM magnetization direction is investigated (see Figure S11 in the Supporting Information).
Furthermore, we investigate the double-stranded DNA with the MPE, and show $MR>P_{FM}$ at the extremely weak SOC (see Section S2, Figure S12 and Figure S13 in the Supporting Information),
which validates the robustness and general applicability of our theoretical framework.

In conclusion, the observation of MR beyond the Jullière model for spin selectivity in chiral molecules has puzzled the community for one decade, and here we propose a solution to this problem by taking into account the MPE.
Due to the MPE, the nonmagnetic-FM interface is shifted, leading to significant modification of the FM polarization and the device MR far beyond the polarization of the FM electrode. Using a standard numerical approach, we show that the MR indeed exceeds the FM polarization in a wide range of model parameters, even for extremely weak SOC.
These results are in good agreement with experimental observations and are expected to end the dispute on the issue of the origin of the CISS.
Ultimately, we also hope that the aforementioned results can be verified using first-principles calculations in the future.\citep{ref_AFM2025,ref_revarxiv_2025}


\vspace{1em}

\noindent ASSOCIATED CONTENT

\noindent \textbf{Supporting Information}

The Supporting Information is available free of charge at: \url{https://github.com/btc-99/Supplemental_Information_of_Julliere_Model}.

Section S1: The Model of $\alpha$-helical Protein; Section S2: The spin polarization and MR of Double-Stranded DNA. (PDF)

\vspace{1em}

\noindent AUTHOR INFORMATION

\noindent \textbf{Corresponding Author}

Qing-Feng Sun -- International Center for Quantum Materials, School of Physics, Peking University, Beijing, 100871, China; Hefei National Laboratory, Hefei 230088, China. Email: sunqf@pku.edu.cn

\noindent \textbf{Authors}

Tian-Yi Zhang -- International Center for Quantum Materials, School of Physics, Peking University, Beijing, 100871, China

Yue Mao -- International Center for Quantum Materials, School of Physics, Peking University, Beijing, 100871, China

Peng-Yi Liu -- International Center for Quantum Materials, School of Physics, Peking University, Beijing, 100871, China

Ai-Min Guo -- Hunan Key Laboratory for Super-microstructure and Ultrafast Process, School of Physics, Central South University, Changsha 410083, China

\noindent \textbf{Author Contributions}

Q.-F.S. designed research; Q.-F.S. supervised the research; T.-Y.Z. performed research; T.-Y.Z. analyzed data; and T.-Y.Z., Y.M., P.-Y.L., A.-M.G., and Q.-F.S. wrote the paper together.

\noindent \textbf{Notes}

The authors declare no competing financial interest.

\begin{acknowledgement}

This work was financially supported by
the National Key R and D Program of China (Grant No. 2024YFA1409002),
National Natural Science Foundation of China (Grant No. 12374034, No. 11921005, and No. 12274466), Quantum Science and Technology-National Science and Technology Major Project (2021ZD0302403), and Hunan Provincial Science Fund for Distinguished Young Scholars (Grant No. 2023JJ10058). We also acknowledge the High-performance Computing Platform of Peking University for providing computational resources.

\end{acknowledgement}

\bibliography{mainbibs}

@ARTICLE{ref1,
   author       = "D. Hsieh and Y. Xia and L. Wray and D. Qian and A. Pal and others",
   year         = "2009",
   journal      = "Science",
   volume       = "323",
   pages        = "919",
   title        = "Observation of unconventional quantum spin textures in topological insulators",
}

@ARTICLE{ref2,
   author       = "U. T. Bornscheuer and G. W. Huisman and R. J. Kazlauskas and S. Lutz and J. C. Moore and K. Robins", 
   year         = "2012",
   journal      = "Nature",
   volume       = "485",
   pages        = "185",
   title        = "Engineering the third wave of biocatalysis",
}

@ARTICLE{ref3,
   author       = "B. Göhler and V. Hamelbeck and T. Z. Markus and M. Kettner and G. F. Hanne and Z. Vager and R. Naaman and H. Zacharias", 
   year         = "2011",
   journal      = "Science",
   volume       = "331",
   pages        = "894",
   title        = "Spin selectivity in electron transmission through self-assembled monolayers of double-stranded {DNA}",
}

@ARTICLE{ref4,
   author       = "K. M. Alam and S. Pramanik", 
   year         = "2015",
   journal      = "Adv. Funct. Mater.",
   volume       = "25",
   pages        = "3210",
   title        = "Spin filtering through single‐wall carbon nanotubes functionalized with single‐stranded {DNA}",
}

@ARTICLE{ref5,
   author       = "R. Naaman and Y. Paltiel and D. H. Waldeck", 
   year         = "2019",
   journal      = "Nat. Rev. Chem.",
   volume       = "3",
   pages        = "250",
   title        = "Chiral molecules and the electron spin",
}

@ARTICLE{ref6,
   author       = "R. Naaman and Y. Paltiel and D. H. Waldeck", 
   year         = "2020",
   journal      = "J. Phys. Chem. Lett.",
   volume       = "11",
   pages        = "3660",
   title        = "Chiral molecules and the spin selectivity effect",
}

@ARTICLE{ref7,
   author       = "R. Naaman and D. H. Waldeck", 
   year         = "2015",
   journal      = "Annu. Rev. Phys. Chem.",
   volume       = "66",
   pages        = "263",
   title        = "Spintronics and chirality: spin selectivity in electron transport through chiral molecules",
}

@ARTICLE{ref8,
   author       = "K. Ray and S. P. Ananthavel and D. H. Waldeck and R. Naaman", 
   year         = "1999",
   journal      = "Science",
   volume       = "283",
   pages        = "814",
   title        = "Asymmetric scattering of polarized electrons by organized organic films of chiral molecules",
}

@ARTICLE{ref9,
   author       = "Y. Xu and W. Mi", 
   year         = "2023",
   journal      = "Mater. Horiz.",
   volume       = "10",
   pages        = "1924",
   title        = "Chiral-induced spin selectivity in biomolecules, hybrid organic–inorganic perovskites and inorganic materials: a comprehensive review on recent progress",
}

@ARTICLE{ref_natmat2024,
   author       = "R. Sun and K. S. Park and A. H. Comstock and A. McConnell and others", 
   year         = "2024",
   journal      = "Nat. Mater.",
   volume       = "23",
   pages        = "782",
   title        = "Inverse chirality-induced spin selectivity effect in chiral assemblies of $\pi$-conjugated polymers",
}

@ARTICLE{ref_sciadv2025,
   author       = "A. Moharana and Y. Kapon and F. Kammerbauer and D. Anthofer and S. Yochelis and H. Shema and E. Gross and M. Kläui and Y. Paltiel and A. Wittmann", 
   year         = "2025",
   journal      = "Sci. Adv.",
   volume       = "11",
   pages        = "1",
   title        = "Chiral-induced unidirectional spin-to-charge conversion",
}

@ARTICLE{ref10,
   author       = "Z. Xie and T. Z. Markus and S. R. Cohen and Z. Vager and R. Gutierrez and R. Naaman", 
   year         = "2011",
   journal      = "Nano Lett.",
   volume       = "11",
   pages        = "4652",
   title        = "Spin specific electron conduction through {DNA} oligomers",
}

@ARTICLE{ref11,
   author       = "D. Mishra and T. Z. Markus and R. Naaman and M. Kettner and B. Göhler and H. Zacharias and N. Friedman and M. Sheves and C. Fontanesi", 
   year         = "2013",
   journal      = "Proc. Natl. Acad. Sci. U.S.A.",
   volume       = "110",
   pages        = "14872",
   title        = "Spin-dependent electron transmission through bacteriorhodopsin embedded in purple membrane",
}

@ARTICLE{ref12,
   author       = "Y.-H. Kim and others", 
   year         = "2021",
   journal      = "Science",
   volume       = "371",
   pages        = "1129",
   title        = "Chiral-induced spin selectivity enables a room-temperature spin light-emitting diode",
}

@ARTICLE{ref13,
   author       = "A.-M. Guo and Q.-F. Sun", 
   year         = "2012",
   journal      = "Phys. Rev. Lett.",
   volume       = "108",
   pages        = "218102",
   title        = "Spin-selective transport of electrons in {DNA} double helix",
}

@ARTICLE{ref14,
   author       = "A.-M. Guo and Q.-F. Sun", 
   year         = "2014",
   journal      = "Proc. Natl. Acad. Sci. U.S.A.",
   volume       = "111",
   pages        = "11658",
   title        = "Spin-dependent electron transport in protein-like single-helical molecules",
}

@ARTICLE{ref15,
   author       = "T.-R. Pan and A.-M. Guo and Q.-F. Sun", 
   year         = "2015",
   journal      = "Phys. Rev. B",
   volume       = "92",
   pages        = "115418",
   title        = "Effect of gate voltage on spin transport along $\alpha$-helical protein",
}

@ARTICLE{refPRL,
   author       = "C. Wang and Z.-R. Liang and X.-F. Chen and A.-M. Guo and G. Ji and Q.-F. Sun and Y. Yan", 
   year         = "2024",
   journal      = "Phys. Rev. Lett.",
   volume       = "133",
   pages        = "108001",
   title        = "Transverse spin selectivity in helical nanofibers prepared without any chiral molecule",
}

@ARTICLE{refPRB_ZTY,
   author       = "T.-Y. Zhang and Y. Mao and A.-M. Guo and Q.-F. Sun", 
   year         = "2025",
   journal      = "Phys. Rev. B",
   volume       = "111",
   pages        = "205417",
   title        = "Dynamical theory of chiral-induced spin selectivity in electron donor--chiral molecule--acceptor systems",
}

@ARTICLE{refJPCL_LPY,
   author       = "P.-Y. Liu and T.-Y. Zhang and Q.-F. Sun", 
   year         = "2025",
   journal      = "J. Phys. Chem. Lett.",
   volume       = "16",
   pages        = "6500",
   title        = "Dynamical simulation of chiral-induced spin-polarization and magnetization",
}

@ARTICLE{refJPCL_liu2025,
   author       = "P.-Y. Liu and T.-Y Zhang and A.-M Guo and Y. Paltiel and Q.-F. Sun", 
   year         = "2025",
   journal      = "J. Phys. Chem. Lett.",
   volume       = "16",
   pages        = "10426",
   title        = "Spin-to-charge conversion modulated by chiral molecules",
}

@article{ref_zhangICISS,
  title={Spin-to-charge conversion driven by inverse chiral-induced spin selectivity},
  author={Zhang, Tian-Yi and Liu, Peng-Yi and Guo, Ai-Min and Sun, Qing-Feng},
  journal={arXiv preprint arXiv:2509.04022},
  year={2025},
  note  = {Submission date: 2025-09-04, URL: \url{https://arxiv.org/abs/2509.04022} (Accessed: 2025-10-27)}
}

@ARTICLE{ref16,
   author       = "S. Alwan and Y. Dubi", 
   year         = "2021",
   journal      = "J. Am. Chem. Soc.",
   volume       = "143",
   pages        = "14235",
   title        = "Spinterface origin for the chirality-induced spin-selectivity effect",
}

@ARTICLE{ref17,
   author       = "H. J. Eckvahl and N. A. Tcyrulnikov and A. Chiesa and J. M. Bradley and R. M. Young and S. Carretta and M. D. Krzyaniak and M. R. Wasielewski", 
   year         = "2023",
   journal      = "Science",
   volume       = "382",
   pages        = "197",
   title        = "Direct observation of chirality-induced spin selectivity in electron donor–acceptor molecules",
}

@ARTICLE{ref_JACS_hole,
   author       = "H. J. Eckvahl and G. Copley and R. M. Young and M. D. Krzyaniak and M. R. Wasielewski", 
   year         = "2024",
   journal      = "J. Am. Chem. Soc.",
   volume       = "146",
   pages        = "24125",
   title        = "Detecting chirality-induced spin selectivity in randomly oriented radical pairs photogenerated by hole transfer",
}

@ARTICLE{ref_PNAS_hole,
   author       = "E. I. Latawiec and A. Chiesa and Y. Qiu and N. A. Tcyrulnikov and R. M. Young and S. Carretta and M. D. Krzyaniak and M. R. Wasielewski", 
   year         = "2025",
   journal      = "Proc. Natl. Acad. Sci. U.S.A.",
   volume       = "122",
   pages        = "e2515120122",
   title        = "Detecting chirality-induced spin selectivity in chromophore-linked {DNA} hairpins using photogenerated radical pairs",
}

@ARTICLE{ref18,
   author       = "D. Nürenberg and H. Zacharias", 
   year         = "2019",
   journal      = "Phys. Chem. Chem. Phys.",
   volume       = "21",
   pages        = "3761",
   title        = "Evaluation of spin-flip scattering in chirality-induced spin selectivity using the {Riccati} equation",
}

@ARTICLE{ref19,
   author       = "T. K. Das and F. Tassinari and R. Naaman and J. Fransson", 
   year         = "2022",
   journal      = "J. Phys. Chem. C",
   volume       = "126",
   pages        = "3257",
   title        = "Temperature-dependent chiral-induced spin selectivity effect: experiments and theory",
}

@ARTICLE{ref20,
   author       = "K. H. Huisman and J.-B. M.-Y. Heinisch and J. M. Thijssen", 
   year         = "2023",
   journal      = "J. Phys. Chem. C",
   volume       = "127",
   pages        = "6900",
   title        = "Chirality-induced spin selectivity ({CISS}) effect: magnetocurrent–voltage characteristics with coulomb interactions i",
}

@ARTICLE{ref_Isr_JF,
   author       = "J. Fransson", 
   year         = "2022",
   journal      = "Isr. J. Chem.",
   volume       = "62",
   pages        = "e202200046",
   title        = "The chiral induced spin selectivity effect what it is, what it is not, and why it matters",
}

@ARTICLE{ref21,
   author       = "P. C. Mondal and N. KantorUriel and S. P. Mathew and F. Tassinari and C. Fontanesi and R. Naaman", 
   year         = "2015",
   journal      = "Adv. Mater.",
   volume       = "27",
   pages        = "1924",
   title        = "Chiral conductive polymers as spin filters",
}

@ARTICLE{ref22,
   author       = "J. M. Abendroth and N. Nakatsuka and M. Ye and D. Kim and E. E. Fullerton and A. M. Andrews and P. S. Weiss", 
   year         = "2017",
   journal      = "ACS Nano",
   volume       = "11",
   pages        = "7516",
   title        = "Analyzing spin selectivity in {DNA}-mediated charge transfer via fluorescence microscopy",
}

@ARTICLE{ref23,
   author       = "H. Aizawa and T. Sato and S. Maki-Yonekura and K. Yonekura and K. Takaba and T. Hamaguchi and T. Minato and H. M. Yamamoto", 
   year         = "2023",
   journal      = "Nat. Commun.",
   volume       = "14",
   pages        = "4530",
   title        = "Enantioselectivity of discretized helical supramolecule consisting of achiral cobalt phthalocyanines via chiral-induced spin selectivity effect",
}

@ARTICLE{ref24,
   author       = "Y. Adhikari and T. Liu and H. Wang and Z. Hua and H. Liu and E. Lochner and P. Schlottmann and B. Yan and J. Zhao and P. Xiong", 
   year         = "2023",
   journal      = "Nat. Commun.",
   volume       = "14",
   pages        = "5163",
   title        = "Interplay of structural chirality, electron spin and topological orbital in chiral molecular spin valves",
}

@ARTICLE{ref25,
   author       = "V. Kiran and S. R. Cohen and R. Naaman", 
   year         = "2017",
   journal      = "J. Chem. Phys.",
   volume       = "146 (9)",
   pages        = "092302",
   title        = "Structure dependent spin selectivity in electron transport through oligopeptides",
}

@ARTICLE{ref26,
   author       = "T. Liu and others", 
   year         = "2020",
   journal      = "ACS Nano",
   volume       = "14",
   pages        = "15983",
   title        = "Linear and nonlinear two-terminal spin-valve effect from chirality-induced spin selectivity",
}

@ARTICLE{ref27,
   author       = "S. Mishra and S. Pirbadian and A. K. Mondal and M. Y. El-Naggar and R. Naaman", 
   year         = "2019",
   journal      = "J. Am. Chem. Soc.",
   volume       = "141",
   pages        = "19198",
   title        = "Spin-dependent electron transport through bacterial cell surface multiheme electron conduits",
}

@ARTICLE{ref28,
   author       = "C. Kulkarni and A. K. Mondal and T. K. Das and G. Grinbom and F. Tassinari and M. F. J. Mabesoone and E. W. Meijer and R. Naaman", 
   year         = "2020",
   journal      = "Adv. Mater.",
   volume       = "32",
   pages        = "1904965",
   title        = "Highly efficient and tunable filtering of electrons' spin by supramolecular chirality of nanofiber‐based materials",
}

@ARTICLE{ref29,
   author       = "P. M. Tedrow and R. Meservey", 
   year         = "1971",
   journal      = "Phys. Rev. Lett.",
   volume       = "26",
   pages        = "192",
   title        = "Spin-dependent tunneling into ferromagnetic nickel",
}

@ARTICLE{ref30,
   author       = "M. Julliere", 
   year         = "1975",
   journal      = "Phys. Lett. A",
   volume       = "54",
   pages        = "225",
   title        = "Tunneling between ferromagnetic films",
}

@ARTICLE{ref31,
   author       = "J. S. Moodera and G. Mathon", 
   year         = "1999",
   journal      = "J. Magn. Magn. Mater.",
   volume       = "200",
   pages        = "248",
   title        = "Spin polarized tunneling in ferromagnetic junctions",
}

@ARTICLE{ref32,
   author       = "U. Huizi-Rayo and J. Gutierrez and J. M. Seco and V. Mujica and I. Diez-Perez and J. M. Ugalde and A. Tercjak and J. Cepeda and E. San Sebastian", 
   year         = "2020",
   journal      = "Nano Lett.",
   volume       = "20",
   pages        = "8476",
   title        = "An ideal spin filter: long-range, high-spin selectivity in chiral helicoidal 3-dimensional metal organic frameworks",
}

@ARTICLE{ref33,
   author       = "H. Lu and others", 
   year         = "2020",
   journal      = "J. Am. Chem. Soc.",
   volume       = "142",
   pages        = "13030",
   title        = "Highly distorted chiral two-dimensional tin iodide perovskites for spin polarized charge transport",
}

@ARTICLE{ref34,
   author       = "H. Al-Bustami and S. Khaldi and O. Shoseyov and S. Yochelis and K. Killi and I. Berg and E. Gross and Y. Paltiel and R. Yerushalmi", 
   year         = "2022",
   journal      = "Nano Lett.",
   volume       = "22",
   pages        = "5022",
   title        = "Atomic and molecular layer deposition of chiral thin films showing up to 99\% spin selective transport",
}

@ARTICLE{ref35,
   author       = "S. Mishra and A. K. Mondal and S. Pal and T. K. Das and E. Z. B. Smolinsky and G. Siligardi and R. Naaman", 
   year         = "2020",
   journal      = "J. Phys. Chem. C",
   volume       = "124",
   pages        = "10776",
   title        = "Length-dependent electron spin polarization in oligopeptides and {DNA}",
}

@ARTICLE{ref36,
   author       = "D. H. Waldeck and R. Naaman and Y. Paltiel", 
   year         = "2021",
   journal      = "APL Mater.",
   volume       = "9",
   pages        = "040902",
   title        = "The spin selectivity effect in chiral materials",
}

@ARTICLE{ref37,
   author       = "S. Alwan and A. Sharoni and Y. Dubi", 
   year         = "2024",
   journal      = "J. Phys. Chem. C",
   volume       = "128",
   pages        = "6438",
   title        = "Role of electrode polarization in the electron transport chirality-induced spin-selectivity effect",
}

@ARTICLE{ref38,
   author       = "T. Liu and P. S. Weiss", 
   year         = "2023",
   journal      = "ACS Nano",
   volume       = "17",
   pages        = "19502",
   title        = "Spin polarization in transport studies of chirality-induced spin selectivity",
}

@ARTICLE{ref39,
   author       = "T.-R. Pan and A.-M. Guo and Q.-F. Sun", 
   year         = "2016",
   journal      = "Phys. Rev. B",
   volume       = "94",
   pages        = "235448",
   title        = "Spin-polarized electron transport through helicene molecular junctions",
}

@ARTICLE{ref40,
   author       = "Q.-F. Sun and X. C. Xie", 
   year         = "2005",
   journal      = "Phys. Rev. B",
   volume       = "71",
   pages        = "155321",
   title        = "Spontaneous spin-polarized current in a nonuniform Rashba interaction system",
}

@ARTICLE{ref_PRB_QD,
   author       = "Q.-F. Sun and J. Wang and H. Guo", 
   year         = "2005",
   journal      = "Phys. Rev. B",
   volume       = "71",
   pages        = "165310",
   title        = "Quantum transport theory for nanostructures with Rashba spin-orbital interaction",
}

@ARTICLE{ref42,
   author       = "Y. Xing and Q. F. Sun and J. Wang", 
   year         = "2008",
   journal      = "Phys. Rev. B",
   volume       = "77",
   pages        = "115346",
   title        = "Influence of dephasing on the quantum {Hall} effect and the spin {Hall} effect",
}

@ARTICLE{ref43,
   author       = "H. Jiang and S. Cheng and Q. F. Sun and X. C. Xie", 
   year         = "2009",
   journal      = "Phys. Rev. Lett.",
   volume       = "103",
   pages        = "036803",
   title        = "Topological insulator: a new quantized spin {Hall} resistance robust to dephasing",
}

@ARTICLE{refNat,
   author       = "J. Schwöbel and Y. Fu, J. Brede and A. Dilullo, G. Hoffmann and S. Klyatskaya and M. Ruben and R. Wiesendanger",
   year         = "2012",
   journal      = "Nat. Commun.",
   volume       = "3",
   pages        = "953",
   title        = "Real-space observation of spin-split molecular orbitals of adsorbed single-molecule magnets",
}

@ARTICLE{refNano,
   author       = "S. L. Kawahara and J. Lagoute and V. Repain and C. Chacon, Y. Girard and S. Rousset and A. Smogunov and C. Barreteau",
   year         = "2012",
   journal      = "Nano Lett.",
   volume       = "12",
   pages        = "4558",
   title        = "Large magnetoresistance through a single nolecule due to a spin-split hybridized orbital",
}

@ARTICLE{refNatNano,
   author       = "M. L. Perrin and C. J. O. Verzijl and C. A. Martin and A. J. Shaikh and R. Eelkema and J. H. Van Esch and J. M. Van Ruitenbeek and J. M. Thijssen and H. S. J. Van Der Zant and D. Dulić",
   year         = "2013",
   journal      = "Nature Nanotech.",
   volume       = "8",
   pages        = "282",
   title        = "Large tunable image-charge effects in single-molecule junctions",
}

@ARTICLE{ref44,
   author       = "T. Morita and S. Kimura", 
   year         = "2003",
   journal      = "J. Am. Chem. Soc.",
   volume       = "125",
   pages        = "8732",
   title        = "Long-range electron transfer over 4 nm governed by an inelastic hopping mechanism in self-assembled monolayers of helical peptides",
}

@ARTICLE{ref45,
   author       = "S. S. Skourtis and I. A. Balabin and T. Kawatsu and D. N. Beratan", 
   year         = "2005",
   journal      = "Proc. Natl. Acad. Sci. U.S.A.",
   volume       = "102",
   pages        = "3552",
   title        = "Protein dynamics and electron transfer: Electronic decoherence and non-{Condon} effects",
}

@ARTICLE{ref46,
   author       = "M. Cordes and B. Giese", 
   year         = "2009",
   journal      = "Chem. Soc. Rev.",
   volume       = "38",
   pages        = "892",
   title        = "Electron transfer in peptides and proteins",
}

@ARTICLE{ref_XYX_PRB,
   author       = "Y. Xing and Q.-F. Sun and J. Wang", 
   year         = "2007",
   journal      = "Phys. Rev. B",
   volume       = "75",
   pages        = "075324",
   title        = "Symmetry and transport property of spin current induced spin-Hall effect",
}

@ARTICLE{ref47,
   author       = "Y. J. Yan and H. Zhang", 
   year         = "2002",
   journal      = "J. Theor. Comput. Chem.",
   volume       = "01",
   pages        = "225",
   title        = "Toward the mechanism of long-range charge transfer in {DNA}: Reduced density matrix approach",
}

@ARTICLE{ref48,
   author       = "G. A. Steele and F. Pei and E. A. Laird and J. M. Jol and H. B. Meerwaldt and L. P. Kouwenhoven", 
   year         = "2013",
   journal      = "Nat. Commun.",
   volume       = "4",
   pages        = "1573",
   title        = "Large spin-orbit coupling in carbon nanotubes",
}

@ARTICLE{ref49,
   author       = "H. Lu and J. Wang and C. Xiao and X. Pan and X. Chen and R. Brunecky and J. J. Berry and K. Zhu and M. C. Beard and Z. V. Vardeny", 
   year         = "2019",
   journal      = "Sci. Adv.",
   volume       = "5",
   pages        = "eaay0571",
   title        = "Spin-dependent charge transport through {2D} chiral hybrid lead-iodide perovskites",
}

@BOOK{ref50,
   author       = "S. Datta", 
   title        = "Electronic Transport in Mesoscopic Systems",
   publisher    = "Cambridge University Press",
   edition      = "First",
   year         = "1995",
}

@ARTICLE{ref_AFM2025,
   author       = "J. Wang and X. Yang and Z. Yang and J. Lu and P. Ho and W. Wang and Y. S. Ang and Z. Cheng and Shibo Fang", 
   year         = "2025",
   journal      = "Adv. Funct. Materials",
   volume       = "202505145",
   title        = "Pentagonal 2D altermagnets: Material screening and altermagnetic tunneling junction device application",
}

@ARTICLE{ref_revarxiv_2025,
  title={Unconventional tunnel magnetoresistance scaling with altermagnets},
  author={Yang, Zongmeng and Yang, Xingyue and Wang, Jianhua and Li, Qiang and Peng, Rui and Lee, Ching Hua and Ang, Lay Kee and Lu, Jing and Ang, Yee Sin and Fang, Shibo},
  journal={arXiv preprint arXiv:2505.17192},
  year={2025},
  note = {Submission date: 2025-05-22, URL: \url{https://arxiv.org/abs/2505.17192} (Accessed: 2025-10-27)}
}

@ARTICLE{ref_revJCP_2023,
   author       = "Y. Kapon and Q. Zhu and S. Yochelis and R. Naaman and R. Gutierrez and G. Cuniberti and Y. Paltiel and V. Mujica", 
   year         = "2023",
   journal      = "J. Chem. Phys.",
   volume       = "159 (22)",
   pages        = "224702",
   title        = "Probing chiral discrimination in biological systems using atomic force microscopy: The role of van der Waals and exchange interactions",
}

\end{document}